\RequirePackage{lineno} 
\documentclass[preprint2]{emulateapj}

\begin{document}

\title{Gamma-ray Emission from Two Blazars \\Behind the Galactic Plane: B2013+370 \& B2023+336}

\author{E. Kara\altaffilmark{1,2}, M. Errando\altaffilmark{1}, W. Max-Moerbeck\altaffilmark{3}, E. Aliu\altaffilmark{1}, M. B\"{o}ttcher\altaffilmark{4}, P. Fortin\altaffilmark{5} \\J. P. Halpern\altaffilmark{6}, R. Mukherjee\altaffilmark{1}, A. C. S. Readhead\altaffilmark{3}, J. L. Richards\altaffilmark{3,7}}

\altaffiltext{1}{Department of Physics \& Astronomy, Barnard College, Columbia University, 3009 Broadway, New York, NY 10027, USA}
\email{E-mail correspondence to :  erin@astro.columbia.edu}
\altaffiltext{2}{Institute of Astronomy, Madingley Road, Cambridge CB3 0HA, UK}
\altaffiltext{3}{Cahill Center of Astronomy and Astrophysics, California Institute of Technology, 1200 E California Blvd, Pasadena CA 91125, USA}
\altaffiltext{4}{Astrophysical Institute, Department of Physics and Astronomy, Clippinger 339, Ohio University, Athens, OH 45701, USA}
\altaffiltext{5}{Laboratoire Leprince-Ringuet, Ecole Polytechnique, CNRS / IN2P3, Palaiseau, France}
\altaffiltext{6}{Columbia Astrophysics Laboratory, Columbia University, 550 West 120th Street, New York, NY 10027, USA}
\altaffiltext{7}{now at Department of Physics, Purdue University, 525 Northwestern Ave, West Lafayette, IN 47907, USA}

\begin{abstract}

\object{B2013+370} and \object{B2023+336} are two blazars at low-galactic latitude  
that were previously proposed to be the counterparts for the EGRET unidentified sources \object[3EG J2016+3657]{3EG~J2016+3657} and \object[3EG J2027+3429]{3EG~J2027+3429}. 
Gamma-ray emission associated with the EGRET sources has been detected
by the {\em Fermi} Gamma-ray Space Telescope, and the two sources, \object[1FGL J2015.7+3708]{1FGL~J2015.7+3708} and \object[1FGL J2027.6+3335]{1FGL~J2027.6+3335}, have been classified as
unidentified in the 1-year catalog.
This analysis of the {\em Fermi}-LAT data collected during 31 months reveals that the 1FGL sources are spatially compatible with the blazars, and are significantly variable, supporting the hypothesis of extragalactic origin for the gamma-ray emission. The gamma-ray light curves are compared with 15 GHz radio light curves from the 40-m telescope at the Owens Valley Radio Observatory (OVRO).  Simultaneous variability is seen in both bands for the two blazar candidates. The study is completed with the X-ray analysis of 1FGL~J2015.7+3708 using {\em Swift} observations that were triggered in August 2010 by a {\em Fermi}-detected flare. The resulting spectral energy distribution shows a two-component structure typical of blazars. We also identify a second source in the field of view of 1FGL~J2027.6+3335 with similar characteristics to the known LAT pulsars. This study gives solid evidence favoring blazar counterparts for these two unidentified EGRET and {\em Fermi} sources, supporting the hypothesis that a number of unidentified gamma-ray sources at low galactic latitudes are indeed of extragalactic origin.  

\end{abstract}

\keywords{Galaxies: individual (B2013+370, B2023+336) -- Gamma rays: galaxies }

\section{Introduction}
\label{intro}

Blazars are a subclass of Active Galactic Nuclei (AGN) which host supermassive black holes in their centers and have relativistic jets pointing along the line of sight to the observer. They show broadband non-thermal emission extending from radio frequencies to gamma rays, with a typical two-component spectral energy distribution \citep[SED, see, e.g.][]{blazar-sed}.
In the {\em Fermi} Large Area Telescope ({\em Fermi}-LAT) 1-year catalog \citep[1FGL,][]{1year} 573 out of 1451 (39\%) detected objects are associated with known blazars, making them the most numerous class of gamma-ray sources.  Only a few bright blazars have shown gamma-ray variability correlated with other wavelengths, allowing a firm identification.  Almost all blazar associations reported in the 1FGL catalog are based on positional coincidence of the {\em Fermi}-LAT gamma-ray source with the radio location of a preselected blazar.
Most gamma-ray blazars are detected at high Galactic latitudes, and only 14 are reported at $|b| < 10^{\circ}$ \citep{1year}.
The same deficit of detections at low Galactic latitudes is reported for gamma-ray emitting AGNs, which includes blazars and also misaligned active galaxies. 
The first {\em Fermi}-LAT AGN catalog \citep[1LAC,][]{fermilac} makes 51 associations with low-latitude AGNs from the VLBA calibrator survey, but a simple extrapolation assuming an isotropic distribution of gamma-ray emitting AGNs indicates that {\em Fermi}-LAT should be detecting $\sim 150$ AGNs at $|b| < 10^{\circ}$ \citep{fermilac}.

\begin{deluxetable*}{ccccccccc}

\tabletypesize{\scriptsize}
\tablecaption{Non-1FGL sources included in the model\label{tab:non1FGL}}
\tablewidth{0pt}
\tablehead{
\colhead{Name} & \colhead{RA} & \colhead{DEC} & \colhead{$\sigma_{95\%}$} & \colhead{TS} &\colhead{Fit} & \colhead{$\Gamma (E_c)$}  & \colhead{Flux 1-100 GeV} & \colhead{2FGL}\\
& \colhead{[deg]} & \colhead{[deg]} & \colhead{[deg]} & & & & \colhead{[$10^{-9}$ $\mathrm{cm^{-2}}$ $\mathrm{s^{-1}}$]} & \colhead{association}
}
\startdata
J2020.0+4158  & 304.70 & 42.16 & 0.08 & 232 & PL & 2.94 $\pm$ 0.03  & 3.50 $\pm$ 0.22 & \object[2FGL J2020.0+4159]{J2020.0+4159}	\\
J2022.2+3840 & 305.55 & 38.68 & 0.13 & 117 & PL & 3.13 $\pm$ 0.04 & 2.45 $\pm$ 0.16 & \object[2FGL J2022.8+3843c]{J2022.8+3843c}\footnote{associated with \object{SNR G76.9+1.0} \citep{arzoumanian}} \\
J2017.4+3628  & 304.45 & 36.42 & 0.02& 465 &PL & 2.59 $\pm$ 0.02 & 7.72 $\pm$ 0.28 & \object[2FGL J2018.0+3626]{J2018.0+3626}\\
J2025.1+3342  & 306.26 & 33.70 & 0.05 & 289 &PL & 2.94 $\pm$ 0.03 & 3.25 $\pm$ 0.14 & \object[2FGL J2025.1+3341]{J2025.1+3341}\\
J2028.3+3333  & 307.08 & 33.55 & 0.03 & 1112&PL Exp & 0.97 $\pm$ 0.05 & 10.74 $\pm$ 0.38 & \object[2FGL J2028.3+3332]{J2028.3+3332} \\
& & & & & &  ($1370 \pm 31$\,MeV) 
\enddata
\end{deluxetable*}

This spatial anisotropy in the distribution of gamma-ray blazars and AGNs is mainly 
caused by a lack of extragalactic source catalogs covering low Galactic latitudes. 
Catalogs of candidate gamma-ray blazars are generally built by selecting compact flat-spectrum radio sources and identifying and classifying them as blazars through optical spectroscopy \citep[see, e.g.][]{cgrabs}. Close to the Galactic plane, diffuse radio emission and confusion with local radio sources make candidate blazars difficult to select. Heavy optical extinction due to interstellar dust \citep{schlegel} further complicates AGN studies at low Galactic latitudes. The three pre-selected catalogs used to find blazar associations for 1FGL sources avoid the Galactic plane due to these observational difficulties \citep{crates,cgrabs,bzcat}. Therefore, the deficit of gamma-ray blazars at low Galactic latitudes is mainly an artifact of the association procedure caused by our limited observational knowledge of blazars behind the Galactic plane. In fact, unbiased gamma-ray surveys like the {\em Fermi} 1-year catalog might be the best tools to identify the population of blazars at low Galactic latitudes that are hardly visible at other wavelengths. 

\begin{deluxetable*}{cccccccc}[b]
\tabletypesize{\scriptsize}
\tablecaption{Analysis results for the gamma-ray sources studied in this work\label{tab:sources}}
\tablewidth{0pt}
\tablehead{
\colhead{Name} & \colhead{RA} & \colhead{DEC} & \colhead{$\sigma_{95\%}$} & \colhead{TS} &
\colhead{Fit} & \colhead{$\Gamma (E_{\rm{c}})$} & \colhead{Flux 1-100 GeV} \\
& \colhead{[deg]} &  \colhead{[deg]} &  \colhead{[deg]}  & & & & \colhead{ [$10^{-9}$ $\mathrm{cm^{-2}}$ $\mathrm{s^{-1}}$] } 
}
\startdata
1FGL~J2015.7+3708  & 303.89 & 37.17 & 0.03& 808 &PL & $2.57 \pm 0.02$ & $9.26 \pm 0.26$\\
J2025.1+3342  & 306.26 & 33.70 & 0.05 & 289 &PL & $2.94 \pm 0.03$ & $3.25 \pm 0.14$\\
J2028.3+3333  & 307.08 & 33.55 & 0.03 & 1112&PL Exp & $0.97 \pm 0.05$ ($1.37 \pm 0.3$\,GeV) & $10.74 \pm 0.38$
\enddata
\end{deluxetable*}

The third EGRET Catalog \citep{3eg} listed 66 high-confidence associations with blazars, but none of them is at $|b| < 10^{\circ}$. Of the 80 reported sources at low Galactic latitudes, 74 were left unidentified. A re-evaluation of the correlation of flat-spectrum radio sources with northern EGRET sources suggested blazar identifications for 5 low-latitude unidentified sources \citep{2023redshift}. Later detailed studies on individual EGRET unidentified sources suggested blazar counterparts for 3EG~J2016+3657 and 3EG~J2027+3429. In this paper we establish firm blazar associations for these two unidentified gamma-ray sources.

The blazar B2013+370 (RA $=20^{\mathrm{h}} 15^{\mathrm{m}} 28.80^{\mathrm{s}}$, DEC $=37^{\circ} 10'  58''$) is located at Galactic latitude $b=1.22^{\circ}$ in the Cygnus region. It was proposed as a counterpart for the EGRET
unidentified source 3EG~J2016+3657 in a comprehensive study of the X-ray
sources in the field followed by optical spectroscopy of the candidate counterparts \citep{mukherjee,halpern}. 
B2013+370 was originally detected as a flat-spectrum radio source \citep{2015discovery,weiler}. It appears compact down to milliarcsecond scales \citep{lee}. There is no spectroscopic redshift, although the optical counterpart
is manifest via its variability \citep{halpern}. \citet{bzcat} classify B2013+370 as a blazar of uncertain type, and \citet{veron} list it as a presumed BL Lac object.
Very High Energy (VHE) gamma-ray observations by the Whipple 10-m telescope led to a 99\% confidence level (c.l.) flux upper limit of $F(E>350\,\mathrm{GeV}) < 2.0 \times 10^{-11} \,\mathrm{cm}^{-2}\,\mathrm{s}^{-1}$ assuming a photon index of $2.5$ \citep{whipple}.

B2023+336 (RA $=20^{\mathrm{h}} 25^{\mathrm{m}} 10.84^{\mathrm{s}}$, DEC $=33^{\circ} 43'  00''$), another blazar in the Cygnus region located at $b=-2.37^{\circ}$, was also proposed as counterpart for an EGRET unidentified source. 
\citet{2023redshift} first suggested the association of 3EG~J2027+3429 with B2023+336 after re-evaluating  the correlation between EGRET unidentified sources and flat-spectrum radio sources. A later study supported the association by reporting significant flux variability from the blazar in the X-ray band \citep{sguera} . Previously, other studies had suggested a young pulsar as counterpart for the EGRET source given its positional coincidence with an OB association \citep{2023pulsar2,2023pulsar1}.
At lower energies, B2023+336 has been detected as a compact flat-spectrum radio source \citep{gb3,lee}. \citet{2023redshift} obtained a spectrum of the optical counterpart coincident with the radio position.
The authors classified B2023+336 as a BL Lac object and derived a redshift of $z=0.219$ based on H spectral lines. However, the signal-to-noise ratio of the optical spectrum is quite low due to heavy optical extinction, and the obtained redshift should be considered as tentative.

In the {\em Fermi} 1-year catalog, the blazars B2013+370 and B2023+336 are respectively found in the vicinity of the unidentified sources 1FGL~J2015.7+3708 and 1FGL~J2027.6+3335 \citep{1year}.

In this paper we determine, discuss, and establish firm blazar associations for the gamma-ray sources 3EG~J2016+3657 and 3EG~J2027+3429 based on new gamma-ray, X-ray, optical and radio data.

\section{Gamma-ray observations}
\label{gamma}

The LAT is the main instrument on-board the {\em Fermi} Gamma-ray Space Telescope. It is a multi-purpose telescope that surveys the gamma-ray sky between 100 MeV and 300 GeV, including the largely unexplored energy window between 10-100 GeV. 

We selected photons from the first 31 months of the {\em Fermi} mission (Aug 2008 - Feb 2011). The analysis was performed using the {\em Fermi}-LAT \texttt{ScienceTools} software package (version v9r18).  Diffuse class photons \cite[Pass6 V3 IRF,][]{LATinstrument} with $E > 300$ MeV were selected in a $17^{\circ} \times 17^{\circ}$ region around B2013+370, a Region of Interest (ROI) which also contains B2023+336.  

\begin{figure}[]
\epsscale{1.12}
\plotone{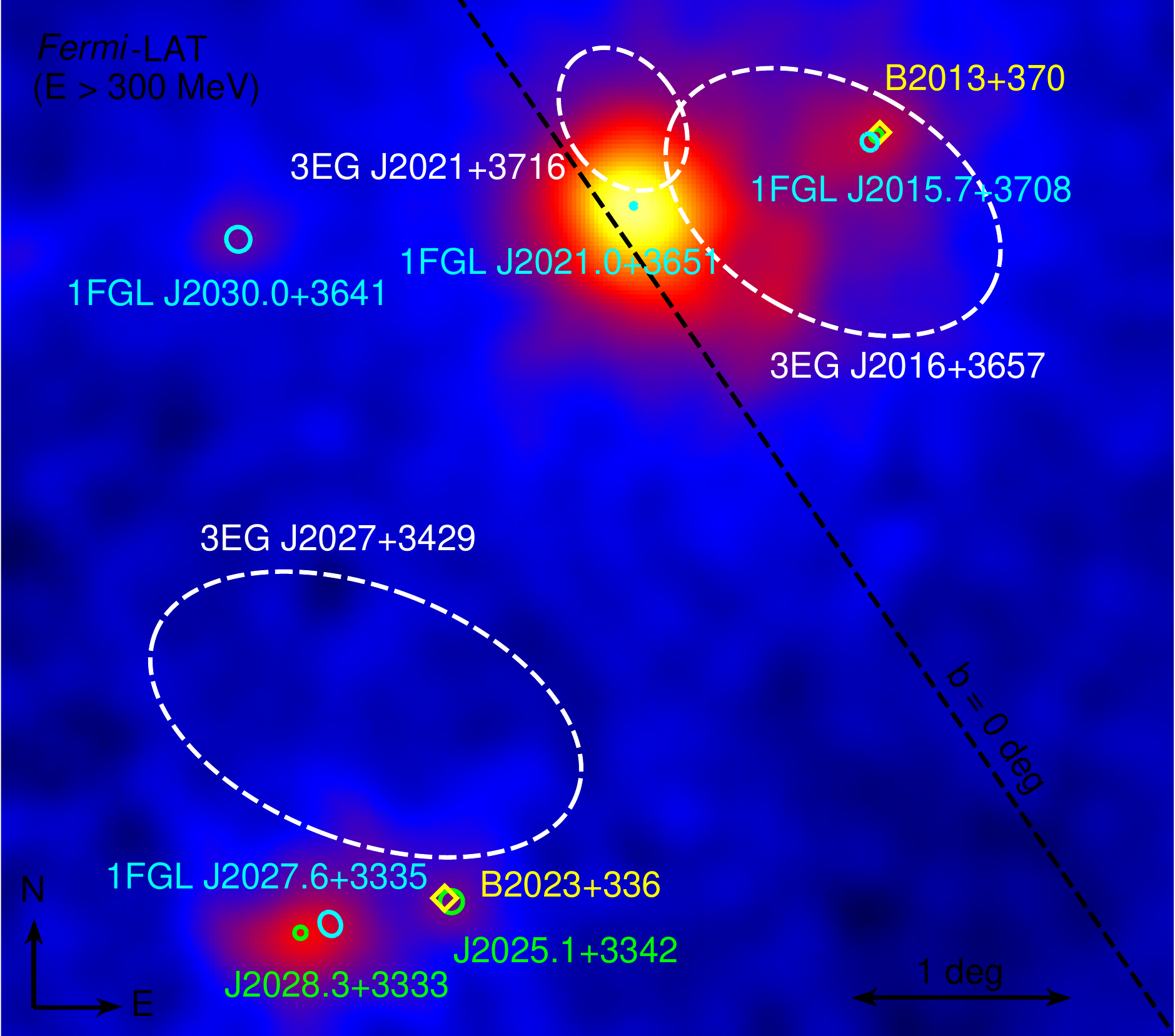}
\hspace{60mm}
\plotone{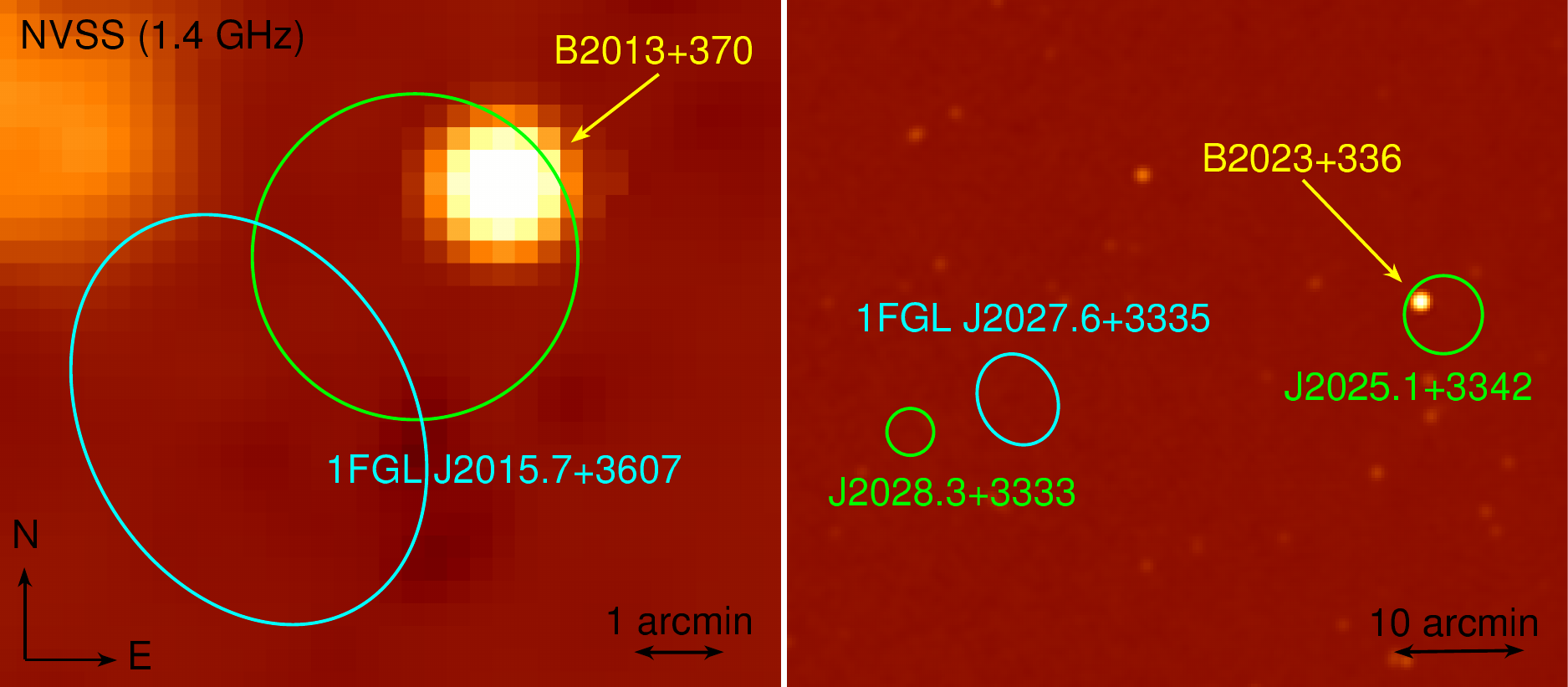}
\caption{Top: {\em Fermi}-LAT counts map ($E>300$ MeV) covering B2013+370 and B2023+336 where the galactic and diffuse gamma-ray emission have been subtracted.  Bottom: Close-up NVSS counts maps \citep[1.4 GHz,][]{NVSS} of B2013+370 (left) and B2023+336 (right).  White dashed contours indicate 95\% c.l. error ellipses for EGRET sources from \citet{mattox-ell}. The error ellipses for the sources in {\em Fermi} 1-year catalog are shown in cyan.  Green circles indicate the 95\% c.l. error circles derived in this analysis, and yellow markers indicate the catalog positions of the two blazars.  
\label{fov}}
\end{figure}

We generated a model that includes all the sources in the {\em Fermi} 1-year catalog inside the ROI (46 1FGL sources total), and the Galactic and isotropic diffuse gamma-ray emission\footnote{This analysis was completed before the {\em Fermi} 2-year catalog was released, and only information from 1FGL was included in the analysis.}. 
The 1FGL sources were modeled by a simple power-law spectrum of the form: $ \mathrm{d}N/\mathrm{d}E = N_0 \left(E/E_0\right)^{-\Gamma} $ where the photon index $\Gamma$ and the normalization $N_0$ were left free.  The only exception is for the known gamma-ray pulsars in the ROI \citep{pulsar}, which have been described by a power-law with exponential cutoff, $\mathrm{d}N/\mathrm{d}E = N_0 \left(E/E_0\right)^{-\Gamma} \mathrm{e}^{-E/E_c} $, where the cutoff energy, $E_c$, was also left free.  Known gamma-ray sources outside the ROI but within $5^{\circ}$ from the edges were included in the model, but their spectral parameters were fixed to the 1FGL values.
The isotropic diffuse background was modeled using the public spectral template (isotropic\_iem\_v02.txt) with a free normalization\footnote{\url{http://fermi.gsfc.nasa.gov/ssc/}}. Galactic diffuse emission was also modeled by a template released by the {\em Fermi} Collaboration (gll\_iem\_v02.fit)$^8$.  Because the sources of interest lie close to the Galactic plane and the background is therefore quite heterogeneous, we weighted the Galactic diffuse emission by a power-law with free normalization and index, yielding a decrease of 11\% on the normalization and a spectral hardening of 3\% with respect to the public template. These additional degrees of freedom significantly improved the overall likelihood of the fit.

A binned likelihood analysis was performed using the \texttt{gtlike} tool.  Using a residuals map, we identified additional sources not in the 1-year catalog, and included them in the model if they had $TS > 25$, equivalent to a  significance $\sigma \simeq 5$ \citep{mattox}.  The location of the additional sources was refined using \texttt{gtfindsrc}.  Details on these additional sources are included in Table~\ref{tab:non1FGL}\footnote{All additional sources have been reported in the 2FGL catalog}.

To derive the spectra and light curves of the studied sources 
the data were divided in time bins for the light curves and energy bins for the spectra, and the likelihood analysis was performed on each bin. To avoid having too many free parameters, the Galactic diffuse emission was fixed. The photon indices of all sources were also fixed, but their normalizations left free.
The spectra were generated using the \texttt{likeSED} python module\footnote{\url{http://fermi.gsfc.nasa.gov/ssc/data/analysis/user/}}. Spectral points in the spectra were required to have $TS>9$. To keep the maximum number of data points in the light curves, flux points were derived when $TS>1$ as described in \citet{fermi-lbas}. Otherwise 90\% confidence level upper limits were calculated.

In our analysis we find that a single source at the location of 1FGL J2027.6+3335 does not represent the data well, and so it is removed from the model and replaced by two separate point-like sources (see Figure~\ref{fov}). A likelihood ratio test (LRT) confirms that the data favors the 2-source hypothesis.  We compared a model with only the {\em Fermi} 1-year catalog source, 1FGL J2027.6+3335, to a model excluding the 1FGL source and including two sources instead: J2025.1+3342 and J2028.3+3333.  The LRT favors the 2-source hypothesis with $\chi^2/$dof = $282/7$, corresponding to $\sim16\,\sigma$.

All results were cross-checked by an independent analysis, and are also found in agreement with a previous analysis of the same field of view including 27 months of data \citep{texasProceedings}.  

In the following we describe more detailed results for these two individual gamma-ray sources, as well as 1FGL 2015.7+3708, all of which are also summarized in Table~\ref{tab:sources}. All reported errors are statistical only. The estimated systematic uncertainty on the flux is 5\% at 500\,MeV, increasing to 20\% at 10\,GeV \citep{1year}.

\subsection{1FGL~J2015.7+3708}
In this analysis 1FGL~J2015.7+3708 is detected with $TS = 808$, equivalent to approximately $28\sigma$.  The location of the gamma-ray source was recalculated using 31 months of data to be RA $=20^{\mathrm{h}} 15^{\mathrm{m}} 33.6^{\mathrm{s}}$, DEC $=37^{\circ} 10'  12''$ with 95\% error circle of 
$1\farcm8$ 
that includes B2013+370. The time-averaged spectrum is well described with a power-law with photon index $\Gamma=2.57 \pm 0.02$, and a measured integral flux above 1\,GeV of $\left(9.26 \pm 0.26\right)\times 10^{-9} \mathrm{cm}^{-2} \mathrm{s}^{-1}$. Our result shows a softer spectrum than the one found in EGRET \citep[$\Gamma=2.09\pm0.11$,][]{3eg}.

\subsection{J2025.1+3342}
In this analysis of 31 months of {\em Fermi}-LAT data we resolve a gamma-ray source, J2025.1+3342, towards the direction of 1FGL~J2027.6+3335. It is located at RA $=20^{\mathrm{h}} 25^{\mathrm{m}} 02.4^{\mathrm{s}}$, DEC $=33^{\circ} 42'  00''$ with 95\% error circle of $3^{\prime}$ that includes the blazar B2023+336. J2025.1+3342 is detected with $TS = 289$, equivalent to approximately $17\sigma$. The energy spectrum is well described by a power-law with a photon index $\Gamma=2.94 \pm 0.03$  and integral flux above 1\,GeV of $\left(3.25 \pm 0.14\right)\times 10^{-9} \mathrm{cm}^{-2} \mathrm{s}^{-1}$.

\subsection{J2028.3+3333}
\begin{figure}
\epsscale{1.15}
\plotone{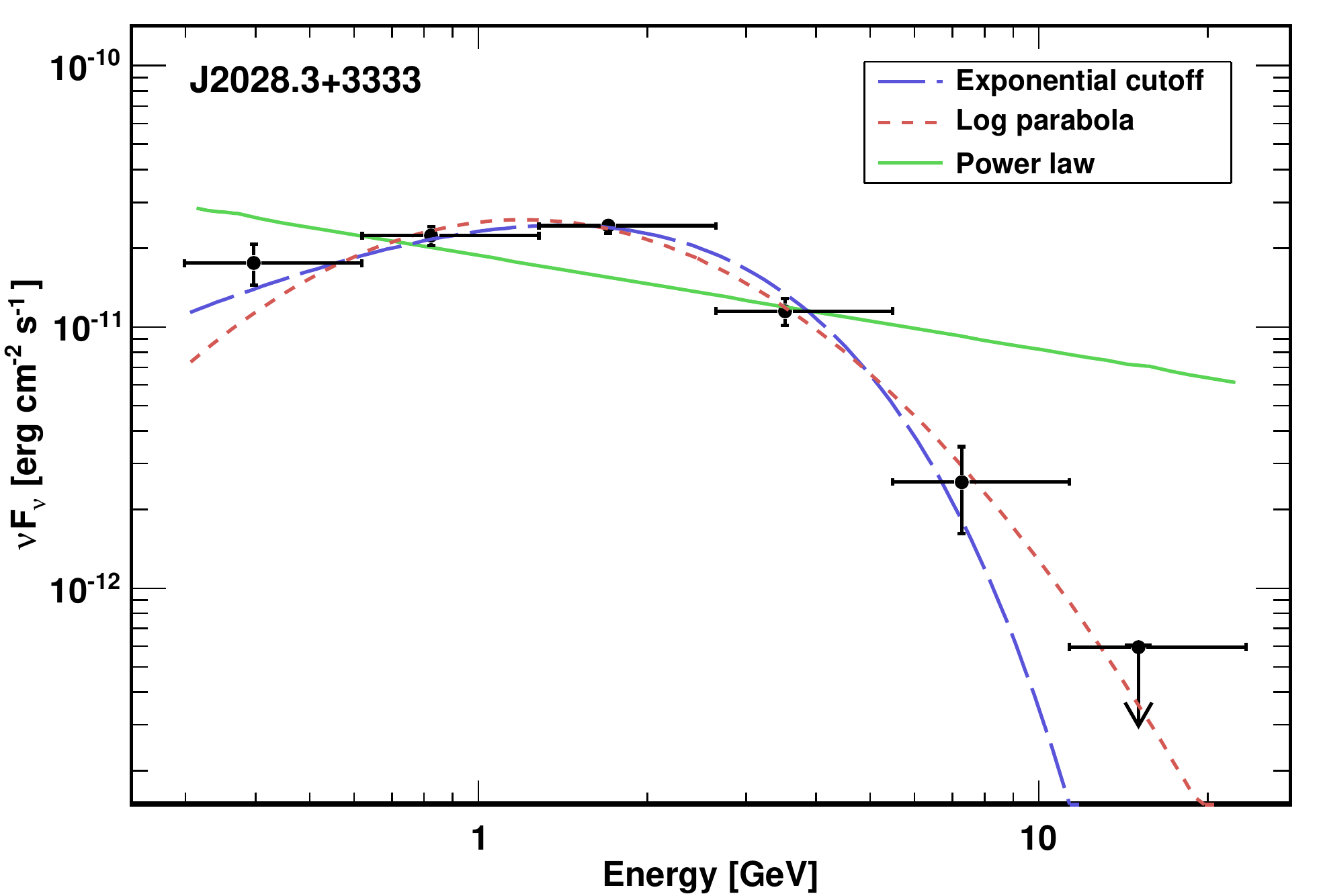}
\caption{Gamma-ray spectrum of J2028.3+3333
. The photon distribution The best-fit descriptions of the photon distribution in the whole energy range   The data was fit to a simple power-law, log-parabola and a power-law with exponential cut-off.  The spectrum is best described by a power-law with exponential cutoff. \label{2028-sed}}
\end{figure}

In addition to J2025.1+3342 we resolve another, brighter gamma-ray emitter, J2028.3+3333, which is likely the primary contributor to 1FGL~J2027.6+3335.   
The second, brighter source, J2028.3+3333, is detected with $TS = 1112$,  or approximately $33\sigma$.  As indicated in \citet{texasProceedings} the energy spectrum is best fit by a power-law with exponential cutoff where $\Gamma = 0.97 \pm 0.05$, $E_{\mathrm{c}}=1.37 \pm 0.03$\,GeV and integral flux above 1\,GeV of $\left(10.74 \pm 0.38\right)\times 10^{-9} \mathrm{cm}^{-2}\mathrm{s}^{-1} $. A LRT comparing a power-law, exponential cutoff and log-parabola spectral fit favors the exponential cutoff with a $\chi^{2}/$dof= $159/1$ ($> 12 \sigma$) with respect to a simple power-law (Figure~\ref{2028-sed}). This source does not exhibit variability in gamma-rays. A $\chi^{2}$ test on a 7-day binned light curve (not shown) gives a probability of the source being variable of $P=3\%$.

\subsection{Comparison with {\it Fermi} 2FGL catalog}
After our gamma-ray analysis was completed, a first version of the {\em Fermi}-LAT 2-year catalog (2FGL) was released \citep{2fgl}. 
The 2FGL catalog, which integrates over 24 months of LAT data, confirms the results obtained for 1FGL~J2015.7+3708, J2025.1+3342, and J2028.3+3333, which are respectively associated with 2FGL~J2015.6+3709, 2FGL~J2025.1+3341, and 2FGL~J2028.3+3332. The location and spectral index agrees within statistical errors for the three sources. Differences up to 20\% in flux normalization are observed in 1FGL~J2015.7+3708 and J2025.1+3342, which could be enhanced by the different integration time and the fact that both sources are variable. Another difference is that 2FGL lists a LogParabola function as  best description for the energy spectrum of the three discussed sources, while our analysis preferred power-law spectra for the two blazars. This difference might be explained by the different energy cuts, as it is observed that the biggest deviation from a power-law spectrum in the 2FGL sources happens to be in the 100-300\,MeV bin, which was excluded from our analysis.
The other reported discrepancies could be caused by multiple analysis differences: models for the Galactic and isotropic diffuse emission, time periods covered (31 months in this study vs. 24 in the 2FGL catalog), event calibrations (P6V3 vs. P7V6), instrument response functions, and energy cuts ($E > 300$\,MeV vs. $E > 100$\,MeV), or versions of the ScienceTools. 
Finally, we note that \object[2FGL J2015.6+3709]{2FGL~J2015.6+3709} and \object[2FGL J2025.1+3341]{2FGL~J2025.1+3341} are respectively associated to the radio counterparts of B2013+370 and B2023+336, while \object[2FGL J2028.3+3332]{2FGL~J2028.3+3332} is left unassociated.
\section{X-ray and optical observations}
\label{x-ray}

\begin{deluxetable}{ccc}[]
\tabletypesize{\scriptsize}
\tablecaption{{\em Swift}-XRT Observation Log\label{tab:swiftobs}}
\tablewidth{0pt}
\tablehead{
\colhead{Observation ID} & \colhead{Start time (UT)} & \colhead{XRT Exposure} \\
& \colhead {[yyyy-mm-dd hh:mm:ss]} & \colhead{[sec]} 
}
\startdata
 00035639001 & 2006-07-12 05:29:01 & 2679.82\\
 00035639002 & 2006-11-12 16:33:01 & 4551.96 \\
 00035639003 & 2006-11-17 04:09:01 & 7449.10\\
00035639004 & 2006-11-24 04:50:01 & 203.65\\
00041471001 & 2010-08-05 18:18:48 & 994.50 \\
00041471002 & 2010-08-06 01:13:38 & 7424.83 \\ 
00041471003 & 2010-08-22 05:35:00 & 4360.92 \\ 
00041471004 & 2010-08-30 01:08:01 & 4216.07 
\enddata
\end{deluxetable}

The {\em Swift} Gamma-Ray Explorer is designed to make prompt observations of Gamma-ray Bursts and afterglows, but also performs pointed observations.
For this analysis, we use data from the {\em Swift} X-ray Telescope (XRT), which is sensitive from 0.2 to 10 keV \citep{swift}.

During August 2010, four {\em Swift} pointed observations of B2013+370 (1FGL~J2015.7+3708) were triggered by a {\em Fermi} detected flare on 2010 August 4\footnote{\url{http://fermisky.blogspot.com/}}. Archival data from observations taken in 2006 were also analyzed. We performed two separate analysis on the 2010 and 2006 observations. Table~\ref{tab:swiftobs} summarizes the {\em Swift}-XRT observations of B2013+370. No {\em Swift} data on B2023+336 are available. 

The XRT data were processed with standard procedures in \texttt{xrtpipeline}, adopting the standard filtering and screening criteria, using FTOOLS in the \texttt{Heasoft} package (v6.10)\footnote{\url{http://heasarc.gsfc.nasa.gov/}}.
For both the 2006 and 2010 sets, we summed the data from the four individual observations.  Due to the low count rate of this weak source ($6.0 \times 10^{-2}$ counts s$^{-1}$ in 2010, and $2.3 \times 10^{-2}$ counts s$^{-1}$ in 2006), we consider only photon counting data, and further select XRT event grades 0-12 \citep{xrtpipeline}.  Pile-up correction was not required.  
From the 2010 and 2006 counts maps, we extracted the source events from a circular region with radius 20 pixels (1 pixel = $2.36''$), and also extracted background events within a 55-pixel radius circle that was offset from the source. 

Ancillary response files were generated with \texttt{xrtmkarf}, accounting for the different extraction regions, vignetting and PSF corrections.  We used current response matrix file v011 available from HEASARC Calibration Database\footnote{\url{http://heasarc.nasa.gov/docs/heasarc/caldb}}.  The spectra were rebinned in order to have at least 20 counts per energy bin, and the spectral fitting were completed using \texttt{XSPEC v12.6.0q}. 

The X-ray spectra were fit by a power-law $F(E)=KE^{-\Gamma}$, absorbed by the Galactic column density, $N^{\rm{Gal}}_{\rm{H}} \sim 1.7 \times 10^{22}$ cm$^{-2}$, which was determined from the data. 
The spectrum from the August 2010 observations resulted in a photon index $\Gamma = 1.77^{+0.22}_{-0.20}$,  and normalization at 1\,keV of $K=1.46^{+0.57}_{-0.39} \times 10^{-3} \: \mathrm{keV}^{-1} \: \mathrm{cm}^{-2} \: \mathrm{s}^{-1}$.  The spectrum from the 2006 observations showed a similar photon index and higher flux $\left(\Gamma = 1.23 ^{+0.55}_{-0.40},\: K = 3.26_{-1.50}^{+0.41} \times 10^{-3} \: \mathrm{keV}^{-1} \: \mathrm{cm}^{-2} \: \mathrm{s}^{-1}\right)$.  

Archival optical observations, imaged on 2004 July 10, were available for the position of B2023+336 from the 2.4m Hiltner Telescope of the MDM Observatory
with a thinned, backside-illuminated SITe CCD having a spatial
scale of 
$0\farcs275$
 per 24$\mu$ pixel.
Three 5 minute exposures in each of the $B, R$, and $I$ filters
were obtained in seeing of $0\farcs8$, and were combined.
An astrometric solution for the images was derived from
the UCAC2 \citep{zacharias}. The optical counterpart
of B2023+336 was clearly detected in each band, at position
R.A.=$20^{\rm h}25^{\rm m}10.\!^{\rm s}85$,
decl.=+$33^{\circ}43^{\prime}00\farcs3$ (J2000.0),
which is consistent with the radio position.
Magnitudes were calibrated with \citet{landolt} standard stars.
Including both statistical and systematic errors, we find
$B=24.12\pm 0.14, R=21.07\pm 0.04$, and $I=19.43\pm 0.04$.

Galactic extinction is a significant factor in the direction of B2023+336, $(l,b) =Ê(73^{\circ}.13, -2^{\circ}.37)$.  The Hydrogen column density \citep[$N_{\mathrm{H}}=6.27 \times 10^{21}$][]{kalberla} was converted to optical extinction, $A_{\mathrm{V}}=2.83$, using the linear relation described in \citet{guver}, and scaled to the selected bandpasses using ratios given in \citet{schlegel}.  The extinction-corrected optical fluxes for the $B$, $R$ and $I$ band can be found in Table~\ref{tab:2027-sed}.

\section{Radio observations}
\label{radio}

Contemporaneous 15~GHz radio light curves were obtained at the
radio positions of the blazars B2013+370 and B2023+336 with the
40-m telescope at the Owens Valley Radio
Observatory (OVRO). Since late 2007, about a year before the
launch of {\em Fermi}, a large-scale, fast-cadence 15~GHz radio
monitoring program has been ongoing using the 40-m telescope
\citep{OVROfermi}.  The OVRO 40-m program began with the 1158
northern ($\delta > -20^{\circ}$) sources from the Candidate
Gamma-ray Blazar Survey~\citep{cgrabs}, and now includes over
1500 sources including all the northern {\em Fermi} 1LAC blazar
associations \citep{fermilac}. The two sources in this analysis
were added to the OVRO 40-m program in August, 2008. Program
sources are each observed twice per week in total intensity with
about 4~mJy (minimum) and 3\% (typical) uncertainty. The absolute
flux density scale is based on the \cite{radioflux} value for
3C~286, 3.44~Jy at 15~GHz, with about 5\% scale uncertainty which
is not reflected in the error bars presented here.

\begin{figure}
\epsscale{1.2}
\plotone{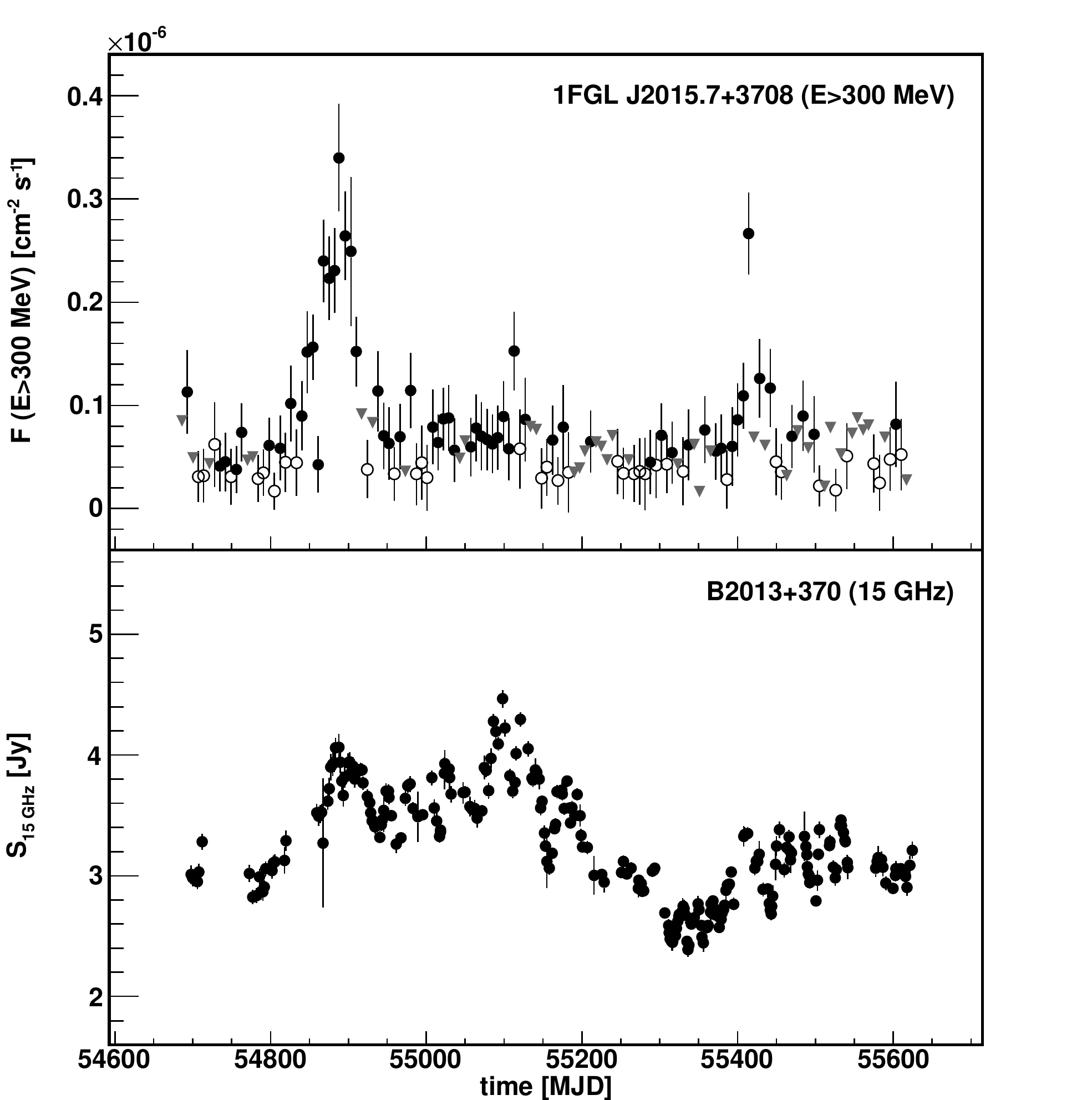}
\caption{ Top: Gamma-ray light curve at $E>300$\,MeV in 7-day bins for 1FGL~J2015.7+3708 (B2013+370). Open circles indicate fluxes for time periods where the source was detected with $TS<4$, and grey triangles represent 90\% c.l. upper limits ($TS<1$).  Bottom:  OVRO radio light curve at 15 GHz for same 31 month period.\label{2015-lc}}
\end{figure}

\begin{figure}
\epsscale{1.16}
\plotone{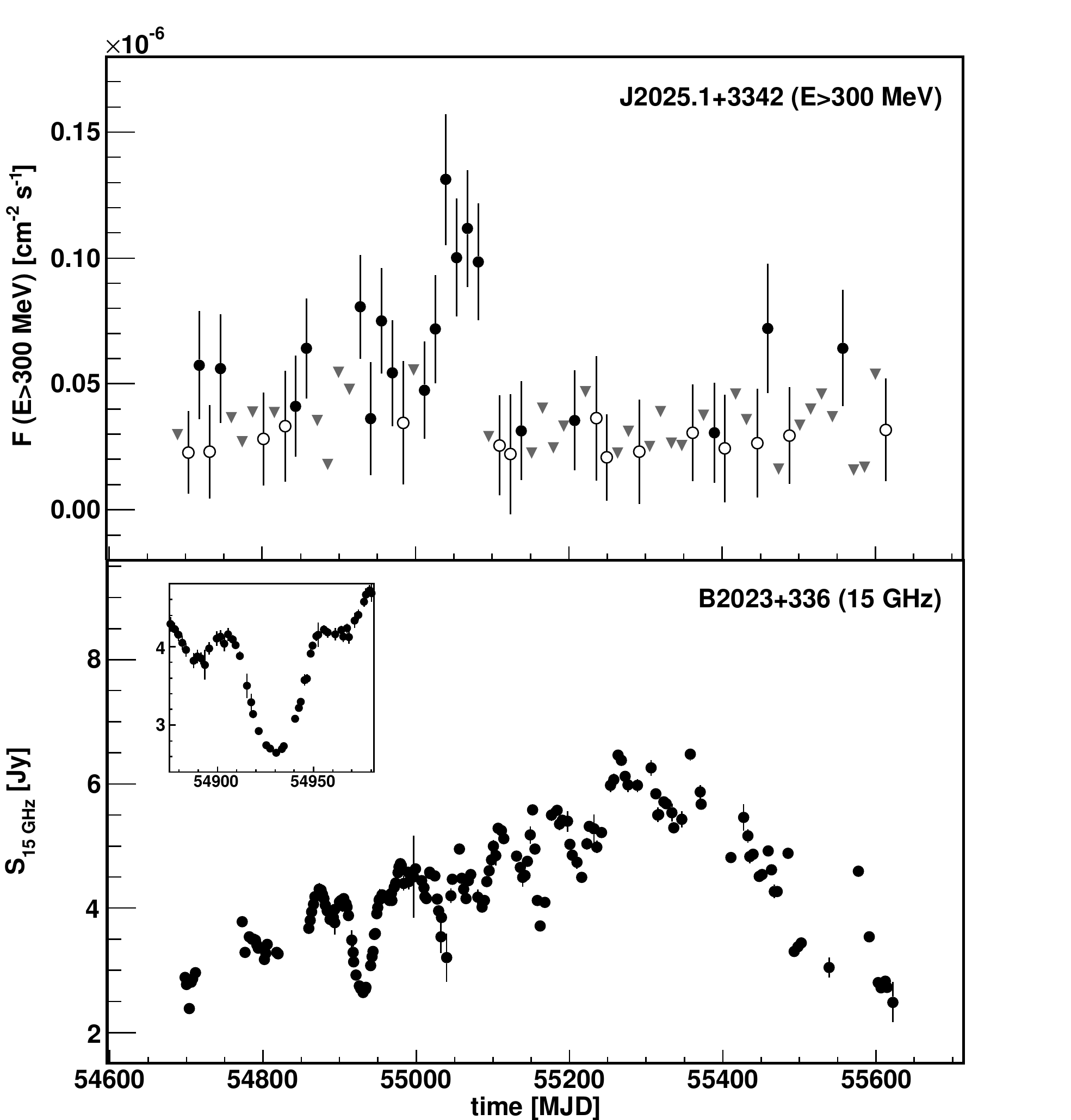}
\caption{Top: Gamma-ray light curve at $E>300$\,MeV in 14-day bins for J2025.1+3342 (B2023+336). Open circles indicate fluxes for time periods where the source was detected with $TS<4$, and grey triangles represent 90\% c.l. upper limits ($TS<1$).  Bottom:  OVRO radio light curve at 15 GHz for same 31 month period. The inset shows a zoom-in of the radio light curve around MJD 54930, when a flux modulation tentatively associated with an extreme scattering event is observed. \label{2025-lc}}
\end{figure}

Radio light curves of B2013+370 and B2023+336 are respectively shown in Figure~\ref{2015-lc} and \ref{2025-lc}, and the time-averaged radio fluxes are reported in Table~\ref{tab:2015-sed} and \ref{tab:2027-sed}.

\section{Location of the gamma-ray source}
\label{location}
A detailed analysis of the {\em Fermi}-LAT data in the direction of B2013+370 and B2023+336 was presented in Section~\ref{gamma}. The resolved gamma-ray sources and their locations with respect to the radio positions of the blazars are shown in Figure~\ref{fov}.

The possible association of the EGRET unidentified source 3EG~J2016+3657 with the blazar B2013+370 was extensively discussed in \citet{mukherjee} and \citet{halpern}. The authors examined X-ray images of the EGRET error box and conducted optical spectroscopy of all candidate counterparts, finding the blazar B2013+370 the only probable counterpart. With an 11-month exposure, the {\em Fermi}-LAT collaboration reported a gamma-ray excess \citep[1FGL~J2015.7+3708,][]{1year} associated with 3EG~J2016+3657. The blazar B2013+370 lies just outside the 1FGL 95\% error ellipse. A slight displacement of the {\em Fermi} reconstructed position could be caused by the high level Galactic diffuse emission in the region and the presence of a bright gamma-ray \object[PSR J2021+3651]{pulsar} only $1.2^{\circ}$ away \cite[PSR~J2021+3651,][]{pulsar}. In the analysis of {\em Fermi}-LAT data presented here, we recalculated the location of the gamma-ray source with 3 times better photon statistics (see Table~\ref{tab:sources}), and found the radio position of B2013+370 compatible with the reconstructed gamma-ray location (Figure~\ref{fov}).  The newly available {\em Fermi} 2-year catalog confirms this association.  Recently, the VERITAS collaboration reported the detection of a VHE gamma-ray source compatible with the nearby pulsar wind nebula \object[CTB 87]{CTB~87} at $E>650$\,GeV \citep{ctb87}. The emission is steady and significantly displaced from B2013+370, suggesting a different origin.

\begin{table}
\begin{center}
\caption{B2023+336 Spectral Points}
\label{tab:2027-sed}
\begin{tabular}{ccc}
\tableline\tableline
Instrument&Frequency & $\nu \mathrm{F}_{\nu}$ \\
&[Hz] & [$10^{-12}$ erg s$^{-1}$ cm$^{-2}$] \\
\tableline
OVRO & 1.5$\times 10^{10}$& $0.428 \pm 0.013$ \\
\hline
MDM & 3.75$\times 10^{14}$& $0.671$ \\
&4.61$\times 10^{14}$ & $0.420$\\
&6.82$\times 10^{14}$ & $0.229$\\
\hline
{\em Fermi}-LAT&1.02$\times 10^{23}$ & $21.2 \pm 2.5$\\
&3.25$\times 10^{23}$ & $6.19\pm 1.13$\\
&1.04$\times 10^{24}$ & $3.31\pm$0.79\\
&3.32$\times 10^{24}$ & $<1.01$ \\
&1.06$\times 10^{25}$ & $<1.29$  \\
\tableline
\end{tabular}
\end{center}
\end{table}

The blazar B2023+336 was proposed as the counterpart for the EGRET unidentified source 3EG~J2027+3429 \citep{2023redshift,sguera}. The {\em Fermi}-LAT 11-month catalog reports an unidentified source (1FGL~J2027.6+3335) outside the EGRET error ellipse.  The position of the 1FGL source is $32'$ away from the radio position of B2023+336. In our analysis presented in Section~\ref{gamma} we resolve a component of 1FGL~J2027.6+3335:  J2025.1+3342, spatially associated with the radio location of B2023+336 (Figure~\ref{fov}).  The recent release of the {\em Fermi} 2-year catalog confirms this association.

\section{Flux variability and correlations}
\label{variability}
Variability in gamma-ray sources allows an identification of the correct counterpart when correlated multi-wavelength variability is found. In addition, stand-alone gamma-ray variability can also support the affiliation of an unidentified source to an intrinsically variable source class \citep[see, e.g.][]{nolan}. Flux variability is one of the main observational characteristics of blazars in the gamma-ray band \citep{fermi-variability}, whereas only a handful of non-AGN sources have shown significant gamma-ray variability: 5 gamma-ray binaries \citep{1year,1018,1259}, V407~Cygni \citep{v407}, and the Crab Nebula \citep{crab-agile,crab-fermi}.

\begin{figure*}[]
\epsscale{1.15}
\plottwo{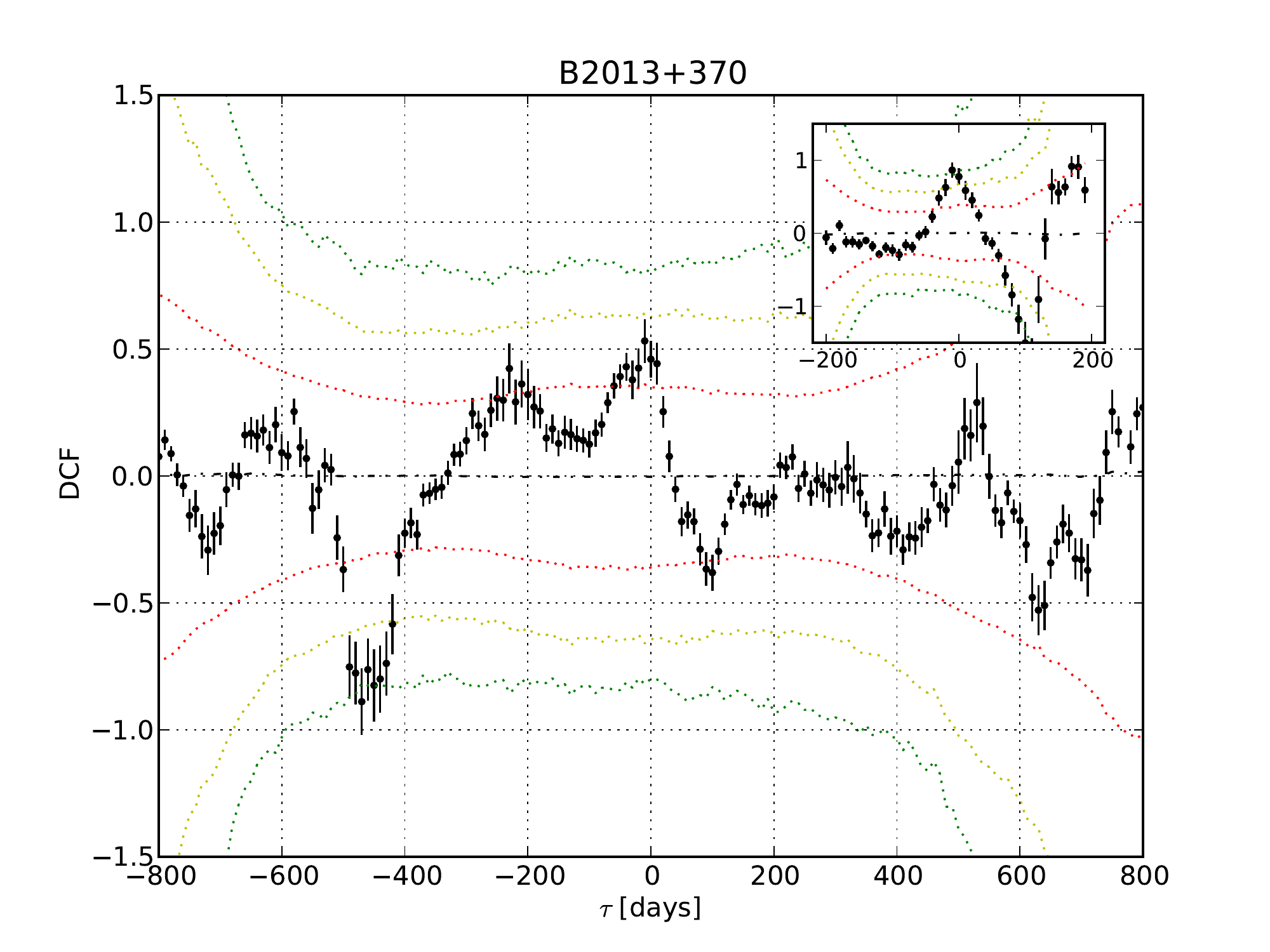}{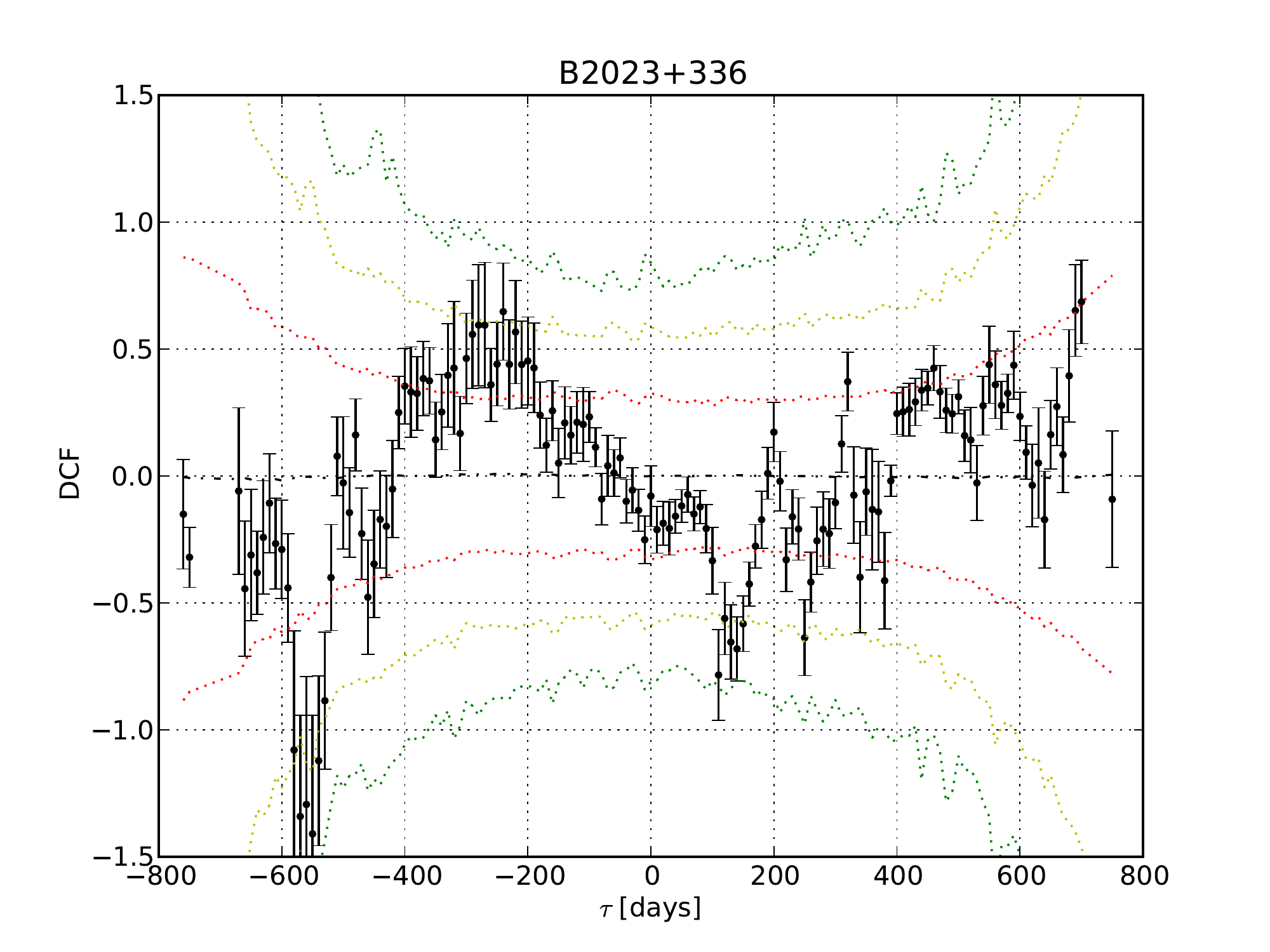}
\caption{Discrete correlation function (DCF) between the gamma-ray and radio light curves for B2013+370 (left) and B2023+336 (right).
Positive time lags indicate that radio activity precedes activity in gamma-rays. The significance of the correlations is illustrated by 1-$\sigma$ red, 2-$\sigma$ yellow and 3-$\sigma$ green dotted lines calculated from simulated uncorrelated data assuming a simple power-law power spectral density with exponent -2.0 for radio and -1.5 for gamma-ray light curves. The inset DCF for B2013+370 during the gamma-ray flare in MJD 54750-55000.
\label{dcfs}}
\end{figure*}

The light curves of the gamma-ray sources associated with B2013+370 and B2023+336 exhibit some degree of variability. We estimated the gamma-ray variability with a $\chi^2$ test on the light curve, including all flux points with $TS>1$ as in \citet{fermi-0agn}. The 7-day binned light curve of 1FGL J2015.7+3708 (B2013+370) shown in Figure~\ref{2015-lc} exhibits clear flaring episodes, and significant evidence of the flux being variable ($P>99.9\%$). J2025.1+3342 (B2023+336) is a weaker gamma-ray emitter. We calculate a probability of its flux being variable of $P=99\%$ based on the 14-day binned light curve shown in Figure~\ref{2025-lc}. This cannot be considered conclusive evidence of the source being variable. However, as noted in \citet{fermilac}, given the same fractional flux variation it is harder to find significant variability in faint sources like J2025.1+3342 than it is for bright sources because of the strong dependence of the variability test on the statistical flux uncertainties.

\begin{table*}
\begin{center}
\caption{B2013+370 Spectral Points\label{tab:2015-sed}}
\begin{tabular}{ccccc}
\tableline\tableline
& \multicolumn{2}{c}{Average} &  \multicolumn{2}{c}{2010 August}\\
\tableline
Instrument & Frequency & $\nu$F$_{\nu}$ & Frequency & $\nu$F$_{\nu}$ \\
 & [Hz] & [$10^{-12}$ erg s$^{-1}$ cm$^{-2}$] & [Hz] & [$10^{-12}$ erg s$^{-1}$ cm$^{-2}$] \\
\tableline
OVRO & 1.5$\times 10^{10}$& $0.485\pm 0.01$ &1.5$\times 10^{10}$ & $0.453 \pm 0.009$ \\
\tableline
{\em Swift}-XRT& 5.11$\times 10^{17}$ & $1.43 \pm 0.10$ & 5.84$\times 10^{17}$& $0.74 \pm 0.09$\\
& 8.42$\times 10^{17}$ & $2.16 \pm 0.15 $& 7.10$\times 10^{17}$ & $2.15 \pm 0.14 $\\
& 1.19$\times 10^{18}$ & $3.05 \pm 0.22 $& 9.66$\times 10^{17}$ & $3.05 \pm 0.24$\\
& 1.51$\times 10^{18}$ & $3.72\pm0.27$ \\
\tableline
{\em Fermi}-LAT &9.81$\times 10^{22}$&$36.1 \pm 10.3 $&1.05$\times 10^{23}$ & $68.8 \pm 13.3$\\
&2.25$\times 10^{23}$&$29.1 \pm $ 3.6 & 3.35$\times 10^{23}$ &$47.7  \pm10.4$\\
&5.16$\times 10^{23}$&$15.8\pm 2.1$ &1.07$\times 10^{24}$ &$24.0 \pm 9.7$\\
 &1.18$\times 10^{23}$&$7.87 \pm 1.54$ &3.42$\times 10^{24}$ &$<10.9$\\
 &2.71$\times 10^{24}$&$3.64 \pm 1.29$ & 1.09$\times 10^{25}$& $<33.2$\\
&6.22$\times 10^{24}$&$<4.53$ & &\\
&1.43$\times 10^{25}$&$<8.35$& &\\
\tableline
\end{tabular}
\end{center}
\end{table*}

A possible correlation between the gamma-ray and radio light curves is studied by constructing a discrete correlation function \citep[DCF,][]{edelson} shown in Figure~\ref{dcfs}. 
The statistical significance of the cross-correlation is investigated with Monte Carlo simulations. It is assumed that the light curves can be described as noise processes with a power-law power spectral density ($\propto f^\beta$). The adopted power-law exponents are -1.5 for gamma-rays \citep{fermi-variability} and -2.0 for radio \citep{chatterjee}. 
Using the recipe prescribed by \citet{timmer}, a pair of independent light curves is generated with a daily sample rate. These light curves are sampled at the same times as the actual radio and gamma-ray light curves and the values are perturbed by adding Gaussian white noise with zero mean and variance that matches the observational errors. The resulting mock light curves are then cross-correlated using the method described in \citet{edelson}. The process is repeated 10000 times, and the results provide an estimate of the distribution of random cross-correlations for each time lag.

The results of the cross-correlation tests are summarized in Figure~\ref{dcfs}, which shows the DCF for the data as black points and the 1-$\sigma$ (red), 2-$\sigma$ (yellow) and 3-$\sigma$ (green) contours for the distribution of random cross-correlations.
Using these values we find no highly significant cross-correlation. The most prominent peaks are located at -10 days (radio lagging) with an 88.0\% significance for B2013+370 and a peak at -240 days with a 96.7\% significance for B2023+336. Additionally, the auto-correlation functions (not shown) were calculated separately for the gamma-ray and radio light curves, without finding any significant peak that could suggest periodic behavior.

Although a statistically significant correlation between the gamma-ray and radio light curves would provide a definite identification, such identifications based on correlated variability have only been established for very bright gamma-ray blazars during short-lived flares \citep[see, e.g.][]{wehrle, iafrate, 3c273}.  
If we only include the prominent gamma-ray flare of B2013+370 in 2009 the value of the cross-correlation peak at -10 days lag clearly increases, as shown in Figure~\ref{dcfs}.
Assessing the statistical significance for very short light curves is difficult because the noise properties used for the statistical test are derived from longer time series and might not be appropriate for short periods of time, especially when these have been selected because of unusual source activity, a prominent gamma-ray flare in this case. The fact that the significance decreases as more data are included indicates that no robust claims about the physical significance of the apparent correlations can be made using time series dominated by a single event.

\begin{figure*}[]
\epsscale{1.1}
\plottwo{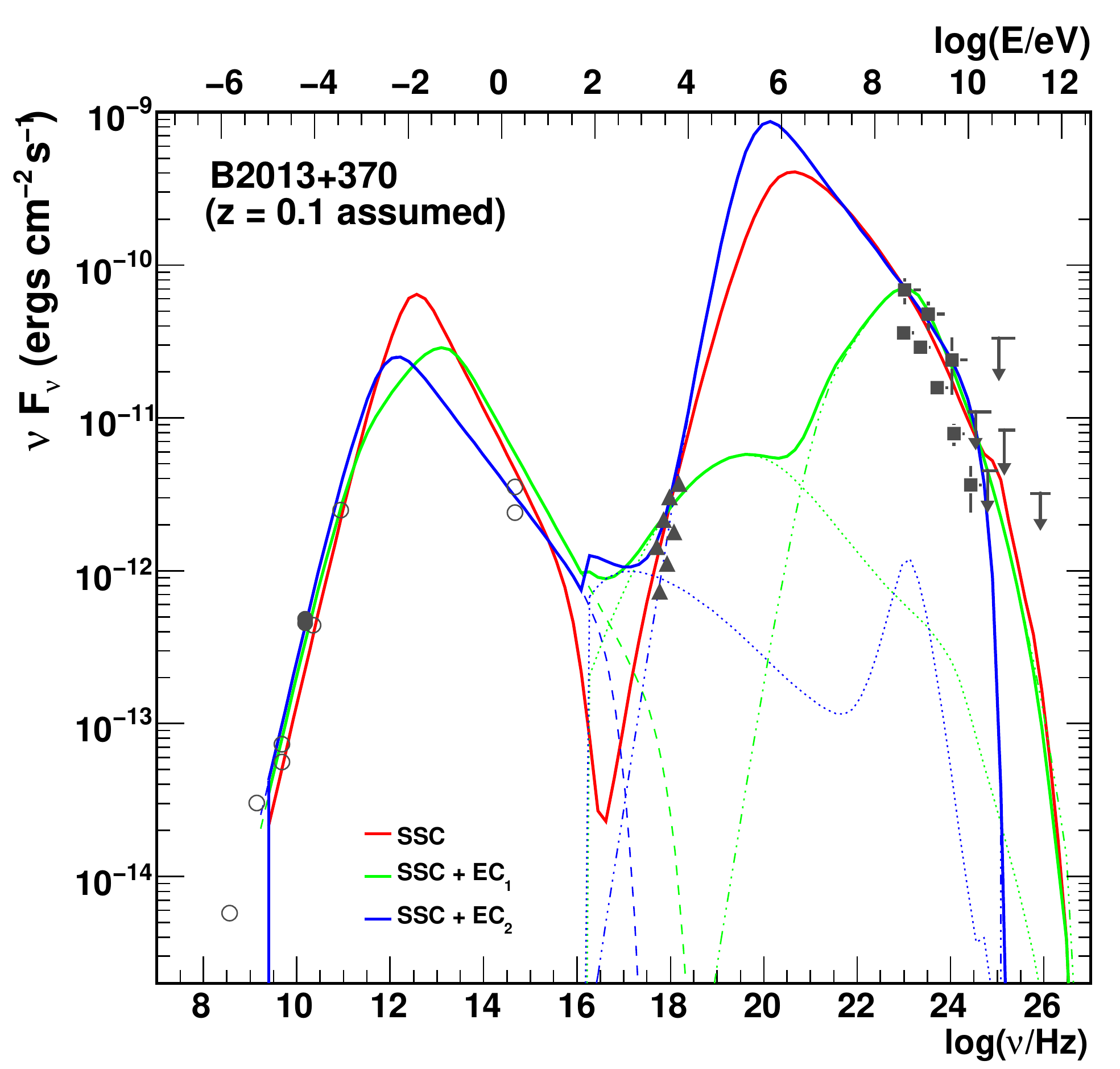}{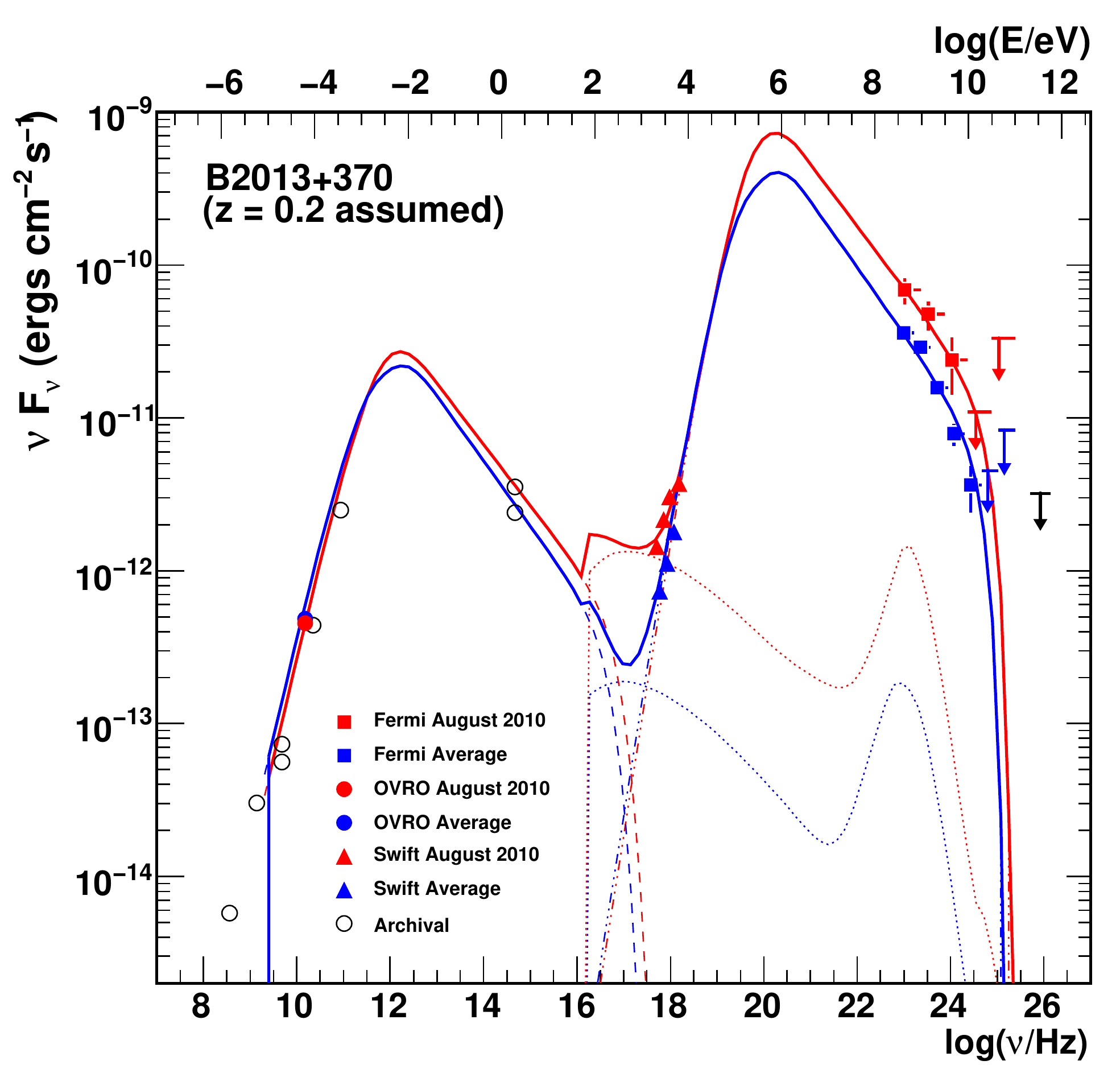}
\caption{Spectral energy distribution for B2013+370. Left: Filled points (this work) correspond to OVRO (15 GHz),  Swift XRT (2-10 keV) and {\em Fermi}-LAT data ($E > 300$\,MeV).  Open circles show archival radio data obtained from NED, {\em R}-band magnitudes from \citet[][]{halpern}, and an $E>350$\,GeV upper limit from \citet[][]{whipple}.  Right: The blue (average) points consist of XRT data from 2006, and an average over the 31 months of Fermi and OVRO data.  The red (August 2010) points consist of the Fermi and OVRO data from the entire month of August 2010, when there was a flare in gamma-rays, which triggered four XRT observations. Solid colored lines indicate the total predicted SED by the different tested models assuming a redshift of $z=0.1$ (left) and $z=0.2$ (right). Dashed lines show the Synchrotron emission, dotted lines the SSC component, and dot-dashed lines the EC. \label{2015-sed}}  
\end{figure*}

\begin{table*}
\begin{center}
\caption{SED modeling parameters\label{tab:2013-model}}
\begin{tabular}{lrccc}
\tableline\tableline
Parameter&Symbol & SSC & SSC+EC$_1$ & SSC+EC$_2$ \\
\tableline
Electron power &$L_e$ [erg s$^{-1}$]&  $4.6\times10^{46}$       &    $4.7\times10^{44}$     &  $1.0\times10^{44}$       \\

Electron low-energy cutoff &$\gamma_{min}$ &     $4.0\times10^{3}$    &     $1.5\times10^{3}$    &     $1.4\times10^{2}$    \\
Electron high-energy cutoff &$\gamma_{max}$&    $3.0\times10^{5}$     &      $3.0\times10^{5}$   &    $2.5\times10^{4}$     \\
Electron injection index &$q_e$                    &      3.2   &   3.1      &  2.8       \\
Equipartition fraction &$L_B/L_e$&$1.58 \times10^{-5}$& $6.49\times10^{-2}$&$1.00$\\
Blob radius &$R_b$ [cm]           &     $7.0\times10^{17}$    &   $6.0\times10^{16}$      &     $1.1\times10^{16}$    \\
Magnetic field &$B$ [G]                   &   0.001      &    0.1     &    1.0     \\
Bulk Lorentz factor &$\Gamma$            &    20     &    15     &     15    \\
Observing jet angle&$\theta_{obs}$ [deg] &    2.87     &  3.82       &   3.82      \\
Redshift (assumed) &$z$                          &   0.1      &   0.1      &   0.1      \\
External radiation energy density &$u_{ext}$ [erg cm$^{-3}$] &    ...     &    $2.0\times10^{-7}$     &    $2.5\times10^{-7}$     \\
External radiation temperature &$T_{ext}$ [K] &    ...     &    $10^3$     &   $10^2$      \\
\tableline
\tableline
\end{tabular}
\end{center}
\end{table*}

Intrinsic long-term flux correlations might be washed out by the presence of extrinsic variability affecting only one of the frequency bands. \citet{spangler} showed that observations of extragalactic sources in the line of sight to B2013+370 and near B2023+336 show heavy scattering due to interstellar plasma turbulence in the direction of the Cygnus OB1 association. This mechanism would produce extrinsic variability in the radio band but not in gamma-rays. Disentangling intrinsic and extrinsic variability in the radio light curves is a difficult problem, but a particular feature on the light curve of B2023+336 around MJD 54930 strongly suggests the possibility of an extreme scattering event \citep{fiedler}. Although extreme scattering events have only been detected at lower radio frequencies, the location of the source at low Galactic latitude and the characteristic symmetric flux decrease expected when an interstellar plasma lens moves through the line of sight, with the small flux increases at both ends of the event being caused by focusing of radio waves at the edges of the plasma cloud, strongly support this hypothesis \citep{clegg}. To explore the effect of excluding the candidate extreme scattering event from the radio light curve of B2023+336, we use the same parameters as for the original analysis but exclude the radio data from MJD 54900 through MJD 54975. We find a slightly higher significance for the cross-correlation of 98.8\% at the same time lag. This supports the suggestion that radio-specific extrinsic effects reduce the observable cross-correlation.

\section{Spectral energy distribution}
\label{sed}

The SED for B2013+370 (Figure~\ref{2015-sed}, data points in Table~\ref{tab:2015-sed}) shows a two component structure characteristic of blazars. 
The relatively hard X-ray spectral index ($\alpha_X=1.2-1.8$) and the dominance of the gamma-ray power output are typically observed in low-frequency-peaked BL Lacs (LBLs) or in flat spectrum radio quasars \citep[\mbox{FSRQs},][]{fossati}. 

We attempted to describe the multi-wavelength SED of B2013+370 with the one-zone leptonic emission model described in \citet[][]{bottcher-chiang} and \citet{wcom}. The model-predicted SEDs are shown in Figure~\ref{2015-sed}, and the assumed model parameters are summarized in Table~\ref{tab:2013-model}. The redshift of the source was assumed to be 0.1 throughout the modeling process.
A pure synchrotron self-Compton model \citep[SSC, e.g.][]{maraschi,bloom} gives a poor description of the SED and requires a very low magnetic field, resulting in a strongly particle dominated jet ($L_B/L_e \sim 10^{-5}$, with $L_B$ and $L_e$ being the magnetic and particle power). This is not surprising: SSC models typically fail to reproduce the high-energy component of Compton dominated blazars (LBLs, and specially FSRQs) unless the magnetic field is reduced to unphysically low values.
This is addressed in external Compton models \citep[EC, e.g.][]{dermer,sikora} by adding a population of low energy photons external to the jet. FSRQs are often described with leptonic models where SSC is dominating in the X-ray band and EC explains the gamma-ray output \citep[e.g.][]{hartman}.
We explore this scenario by adding an external radiation field with temperature $T_{ext}=1000$\,K. The resulting set of parameters (SSC+EC$_1$) is closer to the typical values encountered in blazar SED modeling, but still needs low sub-equipartition magnetic fields and fails to reproduce the observed X-ray spectral slope. The best fit (SSC+EC$_2$) is achieved with both X-rays and gamma-rays being dominated by EC emission. Parameters in exact equipartition ($L_B/L_e = 1$) can be achieved in this case, requiring a rather low temperature of the external radiation field ($T_{ext}=100$\,K) that could originate from cold dust. 
We note that the relatively low frequencies of the photons produced in the first-order SSC ($\sim 10^{17}$\,Hz) make the second-order SSC (peaking at $\sim 10^{23}$\,Hz) become efficient. The lack of an $E\gtrsim 50$\,GeV detection, where the gamma-ray absorption by extragalactic background light is strongly redshift dependent, makes the assumed $z=0.1$ a non-critical model parameter. Similar best-fit solutions close to equipartition were found assuming $z$ up to 1.5. A similar set of model parameters (assuming $z=0.2$ this time) can also describe the average and 2010 August states shown in Figure~\ref{2015-sed}.

\begin{figure}
\epsscale{1.15}
\plotone{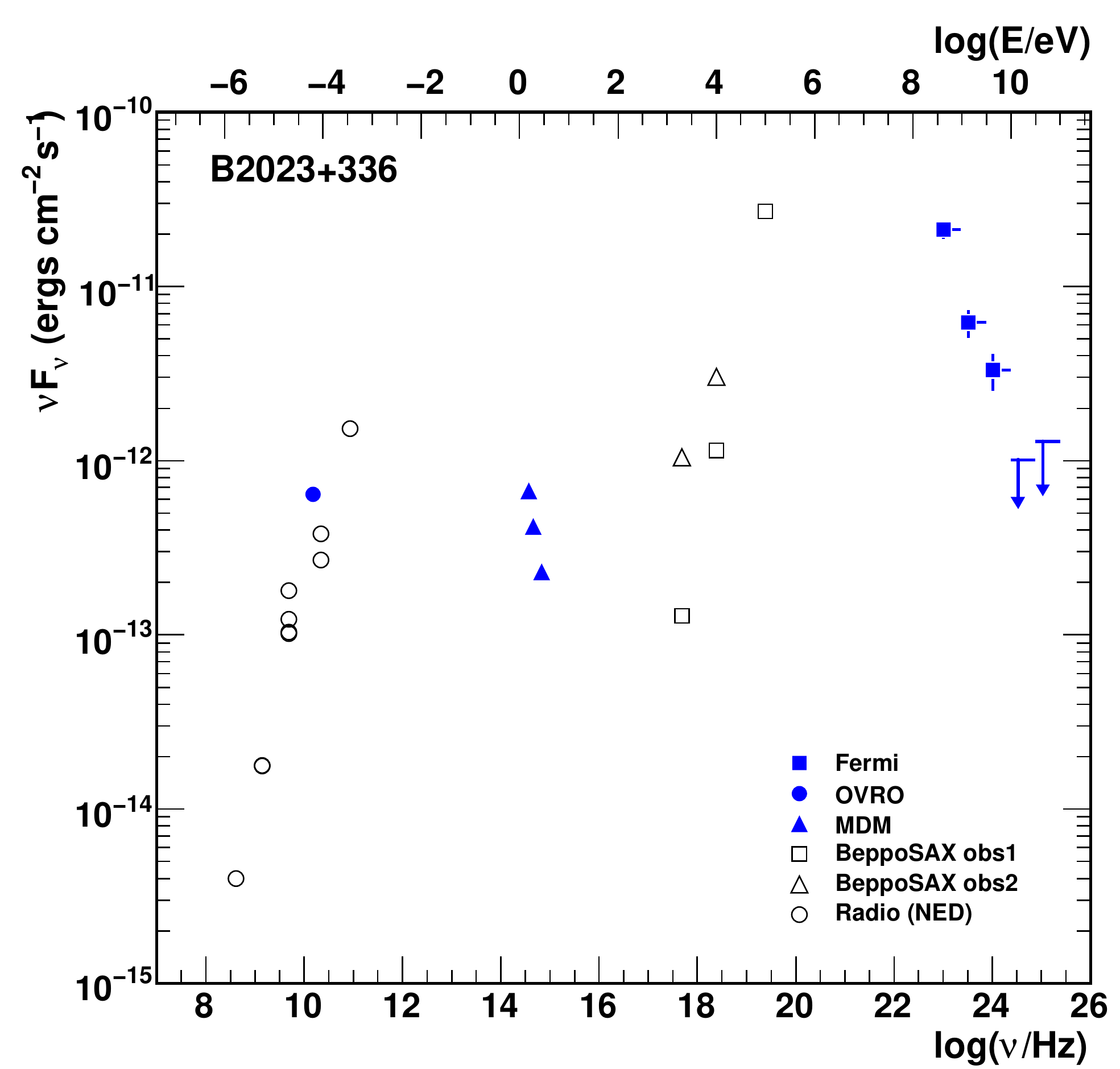}
\caption{Spectral energy distribution for B2023+336 using OVRO (15 GHz), MDM ({\em B}, {\em R}, and {\em I} bands), and {\em Fermi} data ($E > 300$\,MeV).  Open points indicate archival data from NED, and from two observations taken with BeppoSAX (1-100\,keV) from \citet[][]{sguera}. 
 \label{2025-sed}}
\end{figure}

The SED for B2023+336, shown in Figure~\ref{2025-sed}, shows the same blazar-like features discussed for B2013+370: two-component structure, hard X-ray spectrum, and dominant gamma-ray power. However, only archival non-contemporaneous optical and X-ray data are available. Given that B2023+336 is known to be variable at all frequencies, we don not attempt to model this source because of the lack of a simultaneous SED.

\section{Discussion and conclusions}
\label{conclusions}

Spatial association and the observed variability in the gamma-ray and radio bands allow us to establish a firm association between B2013+370 and the previously unidentified gamma-ray source 3EG~J2016+3657 (1FGL~J2015.7+3708), confirming the thesis of \citet{mukherjee} and \citet{halpern}. A compiled SED, adding newly analyzed OVRO, {\em Swift}-XRT and {\em Fermi}-LAT data shows a two component structure that further supports the blazar association. The SED was successfully modeled with a leptonic one-zone model with both X-ray and gamma-ray power dominated by external Comptonization of a low-temperature external radiation field. The gamma-ray dominated SED, hard X-ray spectrum, and preference for EC models point towards B2013+370 being an LBL or an FSRQ.

We also find strong evidence of gamma-ray emission from the blazar B2023+336, that was likely contributing to the unidentified gamma-ray source 3EG~J2027+3429 (1FGL~J2027.6+3335), as hypothesized by \citet{2023redshift} and \citet{sguera}. Our analysis of the available {\em Fermi}-LAT data resolves two independent gamma-ray sources of very different character. J2025.1+3342 is spatially associated with the blazar B2023+336 and shows a hint of variability. The tentative observation of an extreme scattering event in the radio light curve of B2023+336 further supports the hypothesis of it being extragalactic in origin, the flux decrease being caused by a cloud of plasma in the galactic interstellar medium crossing the line of sight.

The other resolved component of 3EG~J2027+3429 (J2028.3+3333) is characterized as a steady gamma-ray emitter with an exponentially cutoff energy spectrum, qualities which are typical of LAT detected pulsars.  This result has recently been confirmed by \cite{pletsch}, who report the discovery of nine previously unknown gamma-ray pulsars using a blind search method. This discovery supports the hypothesis of \citet{2023pulsar2} and \citet{2023pulsar1} that a pulsar could be responsible for the gamma-ray excess 3EG~J2027+3429, or at least part of it.

Optical spectroscopy of the studied objects is complicated due to their low optical luminosity and the extreme dust absorption in the optical band characteristic of low Galactic latitudes. Future observations in large-aperture optical telescopes could establish a definitive spectroscopical identification of both blazars. 

This work has been made possible for the first time because of the continuous, well sampled light curves in the gamma-ray and radio bands obtained over 2.5 years by {\em Fermi} and the fast-cadence OVRO blazar monitoring program. In completing this analysis, we establish firm associations for the blazars B2013+370 and B2023+336 with previously unidentified EGRET and {\em Fermi}-LAT sources. This confirms the gamma-ray band as likely the best tool to identify blazars located behind the Galactic plane, which are heavily absorbed at other wavelengths, and supports the hypothesis that a number unidentified gamma-ray sources at low Galactic latitudes are indeed of extragalactic origin.

\vspace{0.5cm}

\begin{small}
This research is supported by the NASA grant NNX09AU14G and NNX10AP66G, and the US
National Science
Foundation grant Phys-0855627. 
WM acknowledges support from the US Department of State and the Comisi\'{o}n Nacional de
Investigaci\'{o}n Cient\'{i}fica y Tecnol\'{o}gica (CONICYT) in Chile for a Fulbright-CONICYT
scholarship.
The authors thank Diego Tescaro for providing an alternative code to evaluate the discrete correlation function, and Vito Sguera for sharing the X-ray spectra of B2023+336. WM and JR thank Joseph Lazio for a helpful discussion about extreme scattering events.

The OVRO 40~m monitoring program is supported in part by
NASA grants NNX08AW31G and NNG06GG1G and NSF grant AST-0808050.

The \textit{Fermi} LAT Collaboration acknowledges generous ongoing support
from a number of agencies and institutes that have supported both the
development and the operation of the LAT as well as scientific data analysis.
These include the National Aeronautics and Space Administration and the
Department of Energy in the United States, the Commissariat \`a l'Energie Atomique
and the Centre National de la Recherche Scientifique / Institut National de Physique
Nucl\'eaire et de Physique des Particules in France, the Agenzia Spaziale Italiana
and the Istituto Nazionale di Fisica Nucleare in Italy, the Ministry of Education,
Culture, Sports, Science and Technology (MEXT), High Energy Accelerator Research
Organization (KEK) and Japan Aerospace Exploration Agency (JAXA) in Japan, and
the K.~A.~Wallenberg Foundation, the Swedish Research Council and the
Swedish National Space Board in Sweden.

Additional support for science analysis during the operations phase is gratefully acknowledged from the Istituto Nazionale di Astrofisica in Italy and the Centre National d'\'Etudes Spatiales in France.

\end{small}

%\nocite{*}

\clearpage
\newpage

\end{document}